\documentstyle[pre,prb,aps]{revtex}
\begin{document}
% for numbering equations as (section#.equation#):
%
\pagestyle{myheadings}    % Go for customized headings
    \newcommand{\ttt}[1]                % In 'article' only the page
            {                               % number appears in the header.
           \markboth{#1 \hfill}{#1 \hfill} % the page, so I need a new kind
            } 

\ttt{The Camassa-Holm equation as a geodesic flow...}
\title{The Camassa-Holm equation as a geodesic flow on the diffeomorphism
  group.}
\author{Shinar Kouranbaeva$^{\dagger}$}
\address{$^{\dagger}$ Department of Mathematics, University of
  California, Santa Cruz, California 95064}

\date{May 1997; current version July 18, 1998}
\maketitle

\begin{abstract}
Misiolek~\cite{M} has shown that the Camassa-Holm (CH) equation is a 
geodesic flow on the Bott-Virasoro group. In this paper it is shown that 
the Camassa-Holm equation for the case $\kappa =0$ 
is the geodesic spray of the weak 
Riemannian metric on the diffeomorphism group of the line or the circle 
obtained by right translating the $H^1$ inner product over the entire
group. This paper uses the right-trivialisation technique to rigorously 
verify that the Euler-Poincar\'{e} theory for Lie groups can be applied to 
diffeomorphism groups. The observation made in this paper has led to 
physically meaningful generalizations of the 
CH-equation to higher dimensional manifolds (see Refs.~\onlinecite{HMR} and 
~\onlinecite{SH}).
\end{abstract}


\section{Introduction} \label{intro}

Camassa and Holm~\cite{CH,CHH} derived a new completely integrable
dispersive shallow water equation that is bi-Hamiltonian and thus
possesses an infinite number of conservation laws in involution. The
Camassa-Holm (CH) equation is obtained by using an asymptotic expansion
directly in the Hamiltonian for Euler's equations in the shallow water
regime. Below another remarkable property of the CH-equation is shown.
Namely, the CH-equation can be realized as a geodesic equation on a
Riemannian manifold on which the methods of infinite-dimensional
geometry can be applied. Section~\ref{formal} illustrates the main
result by formally applying the Euler-Poincar\'{e} theory for Lie
groups to a continuum mechanical system. The next section verifies the
legitimacy of the application. In addition, Section~\ref{main}
contains independent results on the Riemannian geometry of a $C^1$-
manifold which is also a topological group with $C^1$ right
translation.  Using the right-trivialisation technique, a global
Christoffel map is introduced, and formulas are derived for a spray and a
Levi-Civita connection similar to the finite-dimensional case. 
The method is inspired by the theory of affine connections on
parallelisable manifolds developed by Marsden, Ratiu and 
Raugel~\cite{MRR}.  At the end of section~\ref{main}, a version of the 
Euler-Poincare-Arnold theorem for a diffeomorphism group is verified.
Section~\ref{geo} utilizes the results of section~\ref{main} to
demonstrate 
that the CH-equation is a geodesic flow of the right-invariant metric
on the diffeomorphism group of ${\bf R}$ or of the
circle. Section~\ref{talk} addresses uniqueness and existence issues for 
solutions of the CH-equation. Observations made in this paper have 
$n$-dimensional generalizations to the volume preserving diffeomorphism 
group of a Riemannian manifold which lead to a new class of models for 
mean hydrodynamic motion. See Ref.~\onlinecite{HMR} for application of 
this to numerous fluids models, such as those in geophysics, and see 
Ref. ~\onlinecite{SH} for the development of the geometry and curvature 
of volume preserving diffeomorphism groups with right invariant $H^1$ 
metric. For Riemannian manifolds with boundary, new subgroups of the 
diffeomorphism group have been established which give rise to remarkable 
theorems on the limit of zero viscosity. See Ref.~\onlinecite{HKMRS} for 
a detailed account.

\section {Formal Derivation.} \label{formal}
In this section we illustrate the main result of the paper by 
\textit{formally} applying the pure Euler-Poincar\'{e} theorem on the right-
invariant Lagrangians on Lie groups (see Marsden and Ratiu~\cite{MR} and 
references therein) to the case of the diffeomorphism group of a certain 
Sobolev class $H^s$, $s>\frac{3}{2}$. 
The 
diffeomorphism group is \textit{not} a Lie group (left translation and 
inversion are not smooth, only continuous, whereas right translation is 
smooth), 
and the pure Euler-Poincar\'{e}
theorem strictly does not apply. However, we will demonstrate in the 
following sections that the formal derivation given in this section can be 
rigorously justified using standard trivialisation techniques. 

 Let $M$ be the flat circle $S^1$ or 
the real line $\mathbf{R}$. 
$\mathit{Diff}(M)\equiv \cal{D}$ denotes the diffeomorphism group of $M$ 
of some given Sobolev class. The case $M=S^1$ corresponds to periodic 
boundary
conditions. For the case $M=\mathbf{R}$, the chosen Sobolev space 
automatically imposes appropriate decay 
hypotheses at infinity. Under these boundary conditions, $\mathit{Diff}(M)$ is 
a smooth infinite-dimensional manifold and a topological group relative 
to the induced manifold topology. ${\mathcal X} (M)$ denotes the vector 
fields 
on $M$ of the same differentiability class. Formally, this is the 
\textit{right}  
Lie algebra of $\mathit{Diff}(M)$, e.g., the standard left Lie algebra 
bracket is \textit{minus} the usual Lie bracket for vector fields. For 
$u, v \in {\mathcal X} (M)$ the adjoint action of the Lie algebra on 
itself is given by 
\begin{eqnarray*}
 ad_u v\, =\, [u,v].
\end{eqnarray*}
Consider the $H^1$ inner product on ${\mathcal X} (M)$ and define a 
weak Riemannian metric on the 
whole group $\mathcal{D}$ by right-translation of the given inner 
product on the 
Lie 
algebra. The corresponding norm defines a right-invariant Lagrangian on 
$\mathit{Diff}(M)$ whose restriction to the Lie algebra ${\mathcal X}$ 
is equal to the $H^1$ norm: 
\begin{equation}\label{LagrangianCH}
l(u)\, =\, \frac{1}{2}\int_{M} (u^2+u_x^2) dx.
\end{equation}
Next, one defines $ad^*_u$ the adjoint of $ad_u$ with respect to the  
$H^1$ metric, that is, for $u,v,w \in {\mathcal X}(M)$ 
\begin{eqnarray*}
\left\langle ad^*_u w, v \right\rangle_{H^1}\, =\, \left\langle w, [u,v] 
\right\rangle_{H^1}\, . \end{eqnarray*}

Also, for a function $l:{\mathcal X} (M) \rightarrow\bf{R} $, define the  
functional
derivative $\delta l/\delta u\in {\mathcal X} (M) $ with respect to the 
given metric by
\begin{eqnarray*}
\left\langle \frac{\delta l}{\delta u}, v \right\rangle_{H^1}\, =\, \delta 
l(u)\cdot v \quad \hbox{for} \,\, v \in {\mathcal X} (M)\, .
\end{eqnarray*}

Assuming the existence of $ad^*_u$ for each $u \in {\mathcal X}(M)$, we can 
formally write the Euler-Poincar\'{e} equations
\begin{eqnarray*}
\frac{d}{dt}\frac{\delta l}{\delta u}\, =\, -ad^*_u\frac{\delta l}{\delta 
u}\, .
\end{eqnarray*}
After computing $ad^*_u$ and $\delta l/ \delta u$ (for the 
computations see section~\ref{geo}) the 
Euler-Poincar\'{e} equations yield the Camassa-Holm equation
\begin{equation}
\label{CH}
u_t-u_{xxt}=-3uu_x+2u_xu_{xx}+uu_{xxx}\, .
\end{equation}
The above observation motivates one to develop a theory 
to perform the procedure.

\section{ Riemannian geometry  of partial Lie groups.}
\label{main}
\subsection{Review of definitions}  \label{def}
In what follows, the general structure needed is that of a $C^1$-manifold 
$G$ which is also a topological group in the induced manifold
topology, and it is assumed that only the right translation is $C^1$. We call 
such a group $G$ a partial Lie group. Below we review the 
Riemannian geometry of partial Lie groups, see Spivak~\cite{S}.  
Let $G$ be a manifold equipped with a metric $\langle\cdot,\cdot\rangle$. 
Let 
$\pi_G:TG\rightarrow G$ and $\pi_{TG}:TTG\rightarrow TG$ be the tangent bundle
projections and denote by $V=kerT\pi_G$ the vertical subbundle of $TTG$.
Define the {\it connector} $K:TTG\rightarrow TG$ by
\begin{eqnarray}
  K(TY\cdot X)=\nabla_XY,
\nonumber
\end{eqnarray}
for $X,Y\in {\cal X}(G)$ the Lie algebra of vector fields and $\nabla$
the Levi-Civita connection coming from a metric.

A vector $U\in TTG$ is called {\it horizontal} if $U\in kerK$; $H=kerK$
is a subbundle of $TTG$ called the {\it horizontal subbundle} of the
connection and we have the decomposition $TTG=H\oplus V$ over $TG$ with the
projection $\pi_{TG}$. Then the {\it horizontal lift} of $w\in T_gG$ to
$T_v(TG)$, $v\in T_gG$, is defined as
\begin{eqnarray}
  hor_vw=(T_v\pi_G|H_v)^{-1}(w).
\nonumber
\end{eqnarray}
The horizontal lift operator $hor_v:T_gG\rightarrow H_v$ is an isomorphism
for all  $v\in TG$ and locally,
\begin{eqnarray}
hor_vw=b^i{\partial\over\partial
  x^i}-\Gamma^i_{jk}b^ja^k{\partial\over \partial v^i}\, ,
\nonumber
\end{eqnarray}
where $v=a^i\partial /\partial x^i$ and $w=b^i\partial /\partial x^i$.

The {\it spray} $S:TG\rightarrow TTG$ is by definition the Lagrangian
vector field of the energy function $E(v)=L(v)=\frac{1}{2}\langle 
v,v\rangle $; i.e. \begin{eqnarray}
 {\bf i}_S\omega_L={\bf d}E\, ,
\nonumber
\end{eqnarray} 
where $\omega_L$ is the symplectic form on $TG$, and ${\bf i}_S$ denotes 
the interior product (for 
more 
details refer to {\it Foundations of Mechanics}~\cite{AM}). Locally,
\begin{eqnarray}
 S(v) = a^i{\partial\over\partial x^i} - \Gamma^i_{jk}a^ja^k{\partial
\over\partial v^i}\, ,
\nonumber
\end{eqnarray}
where $v=a^i\partial /\partial x^i$, and notice that by definition, 
we have the useful identity
\begin{equation}
 S(v)=hor_vv\, .
\label{S}
\end{equation} 

 Let $g(t)$ be a smooth curve in G and let $\dot{g}(t)$ be its tangent
vector field. If $Y$ is another vector field, define the {\em covariant
 derivative} of $Y$ along $g(t)$ by
\begin{eqnarray}
 {DY\over dt} = \nabla_{\dot{g}(t)}Y\, .
\nonumber
\end{eqnarray}
If the covariant derivative of $Y$ is zero, $Y$ is said to be parallel
along $g(t)$. It follows from the definition of a connector that
$TY(\dot{g}(t)) \in H$ if and only if $DY/dt = 0$. Locally for a given
curve $g(t)$ this equation becomes a linear system of ordinary differential
equations
\begin{eqnarray}
{dY^i(t)\over dt} + \Gamma^i_{jk}\dot{g}^j(t)Y^k(t) = 0\, .
\nonumber
\end{eqnarray}

A curve $g(t)$ is called the {\em geodesic} of a connection $\nabla$, if
 $\dot{g}(t)$ is parallel along $g(t)$; i.e. if 
\begin{eqnarray}
\nabla_{\dot{g}(t)}\dot{g}(t)=0\, .
\nonumber
\end{eqnarray}
Locally, this is a second-order differential equation
\begin{eqnarray}
\nonumber
 \ddot{g}^i(t) + \Gamma^i_{jk}\dot{g}^j(t)\dot{g}^k(t) = 0\, .
\end{eqnarray}

\subsection{The Levi-Civita connection and the spray for a right-invariant
metric on $G$.} \label{formulas}

Let $G$ be a $C^1$-manifold which is a topological group with $C^1$
right translation.  Assume that $G$ admits a right-invariant metric.
There is a vector bundle isomorphism called the right trivialisation
map
\begin{eqnarray*}
\rho : TG & \longrightarrow & G\times\cal{G} , \\
       v  & \longmapsto &  (g,T_gR_{g^{-1}}\cdot v).  
\end{eqnarray*}
Then $T\rho: TTG\longrightarrow T(G\times \cal{G})$ maps $TTG$
isomorphically onto $TG\times\cal{G}\times\cal{G}$. We can further
trivialize via $\rho\times id$ 
\begin{eqnarray}
TTG \stackrel{(\rho\times id)\circ T\rho}{\longrightarrow}
G\times\cal{G}\times\cal{G}\times\cal{G} .
\nonumber
\end{eqnarray} 
Note the isomorphic image of the vertical subbundle of $TTG$ is equal
to $G\times O\times \cal{G}\times\cal{G}$, the projection being onto
the first and third factors. To keep the base points in the first two
factors, we apply the involution map 
\begin{eqnarray*}
\sigma :G\times\cal{G}\times\cal{G}\times\cal{G}  & \longrightarrow &
G\times\cal{G}\times\cal{G}\times\cal{G} , \\
(g,X,Y,Z) & \longmapsto & (g,Y,X,Z). 
\end{eqnarray*}
Then the image of the vertical bundle $V$ of $TTG$ equals
$G\times\cal{G}\times O\times\cal{G}$ with the projection being on the
first two factors. Therefore, the isomorphism we are working with is
\begin{eqnarray}
TTG \stackrel{\sigma\circ (\rho\times id)\circ T\rho}{\longrightarrow}
G\times\cal{G}\times\cal{G}\times\cal{G} .
\nonumber
\end{eqnarray}

A given metric gives rise to a Levi-Civita 
connection which determines the
horizontal bundle. We wish to express it in
the trivialisation $\rho$, which in turn helps us to find the
spray. Define the continuous $\bf{R}$-bilinear map
$\gamma_g:\cal{G}\times\cal{G}\longrightarrow \cal{G}$ depending
smoothly on $g$ by
\begin{eqnarray}
\label{gamma}
\rho ((\nabla_XY)(g))=(g,d\bar{Y}(g)\cdot
X(g)+\gamma_g(\bar{X}(g),\bar{Y}
(g))),
\\
\nonumber
\hbox{where}\quad \rho(X(g))=(g,\bar{X}(g))\quad\hbox{and}\quad X,Y\in
{\cal X}(G).
\end{eqnarray}
To see a connection in a finite-dimensional case recall that in
coordinates we have 
\begin{eqnarray}
S(X) & = & X^i{\partial\over\partial g^i} - \Gamma^i_{jk}X^jX^k{\partial
\over\partial \dot{g}^i}, \quad \hbox{and}\nonumber \\
(\nabla_XY)^i & = & X^jY^i_{,j} + \Gamma^i_{jk}X^jX^k\, .
\nonumber 
\end{eqnarray}
Define $\gamma = \Gamma^i_{jk}X^jX^k{\partial
\over\partial \dot{g}^i}$, then
\begin{eqnarray}
(\nabla_XY)^i & = & X^jY^i_{,j} + \gamma^i.
\label{finiteCristoffel}
\end{eqnarray}

Let $g(t)$ be a curve in $G$ with $g(0)=g$, $\dot{g}(0)=w$, and 
$v\in T_gG$. There exists a curve $v(t)\in TG$ such that
\begin{equation}
v(0)=v,\quad \pi_G(v(t))=g(t),\quad  {Dv(t)\over dt}=0.
\label{4}
\end{equation}
Therefore, 
$Tv(t)\cdot \dot{g}(t)$ is horizontal; i.e. the tangent vector field 
$dv/dt$ of $v(t)$ is always horizontal. Therefore
  $\dot{v}(0)= hor_vw$, the horizontal lift of $w=\dot{g}(0)\in T_gG$ to
  $T_v(TG)$. If
\begin{eqnarray*}
\rho (v(t)) & = & (g(t),\xi (t)) \qquad \rho (v)= (g,\xi) \\
\rho (\dot{g}(t)) & = & (g(t),\zeta (t)) \qquad 
\rho (w)=(g,\zeta), 
\end{eqnarray*}
then $v(0)=v$ in the trivialisation reads $\xi (0)=\xi$, and the base 
projection 
condition is automatically satisfied. By the definition of a covariant
derivative and the chain rule, we find that 
\begin{eqnarray*}
\rho \left( {Dv\over dt}\right) & = & (g(t),d\xi (t)\cdot \dot{g}(t)
+ \gamma_{g(t)}(\zeta (t),\xi (t))) \\
 & = &  (g(t),{d\xi\over dt}
+ \gamma_{g(t)}(\zeta (t),\xi (t))),
\end{eqnarray*}
so that $(g(t),\xi(t))$ is parallel along $g(t)$ in $G\times \cal{G}$
relative to the push-forward connection by $\rho$ {\it if and only if}
\begin{equation}
{d\xi\over dt}
+ \gamma_{g(t)}(\zeta (t),\xi (t)) = 0, \quad \xi (0)=\xi.
\label{5}
\end{equation}
This equation enables us to compute the horizontal lift of $(g,\zeta)$
to $T_{(g,\xi )}(G\times \cal{G})$ and hence the spray using
(\ref{S}). Let us compute $hor_vw=\dot{v}(0)$ in the trivialisation
given by $\rho$. We have that 
\begin{eqnarray*}
(\sigma \circ (\rho\times id)\circ T\rho )(\dot{v}(0)) & = & \sigma 
\circ (\rho\times id)\lgroup {{d\over dt}\arrowvert}_{t=0}(\rho\circ v)
(t)\rgroup \\
 & = & \sigma \circ (\rho\times id)\lgroup\dot{g}(0),\xi (0),{{d\xi (t)
\over dt}\arrowvert}_{t=0}\rgroup \\
 & = & \sigma (g(0),\zeta (0),\xi (0),-\gamma_{g(0)}(\zeta (0),\xi
 (0))) \\
 & = & (g,\xi ,\zeta ,-\gamma_g(\zeta ,\xi )). 
\end{eqnarray*}
Therefore,
\begin{eqnarray}
hor_{(g,\xi )}(g,\zeta )=(g,\xi ,\zeta ,-\gamma_g(\zeta ,\xi ))
\nonumber
\end{eqnarray}
and the spray of the Levi-Civita connection in its right
trivialisation is given by
\begin{equation}
\bar{S}(g,\xi )=hor_{(g,\xi )}(g,\xi )=(g,\xi ,\xi ,-\gamma_g(\xi ,\xi )).
\label{Sbar}
\end{equation}
Applying $(\rho^{-1}\times id)\circ \sigma^{-1}$ we can express the
right trivialisation of the spray as
\begin{eqnarray}
\bar{S}(g,\xi )= (T_eR_g\cdot\xi ,\xi ,-\gamma_g(\xi ,\xi )),
\label{Sbar2}
\end{eqnarray}
where $X(g)=T_eR_g\cdot\xi$ is the right-invariant vector field
on $G$ associated to a Lie algebra element $\xi$. It follows that
\begin{equation}
S(v)=T\rho^{-1}\circ\bar{S}\circ\rho (v).
\end{equation}
                
Given a vector bundle $E$ over $G$, we shall denote by
$\cal{E}$  the collection of all smooth sections \mbox{$\sigma :G
\rightarrow E$} such that $\pi\circ\sigma = id_G$.
Let \mbox{$E=G\times\cal{G}$}, then the condition  
$\pi\circ\sigma =id_G$ implies that 
\begin{eqnarray*}
\cal{E}\; =\; \{\sigma :G\rightarrow\cal{G}\; |\; \sigma\; \hbox{is
smooth}\}.
\end{eqnarray*}
If $\rho (X(g))=(g,\bar{X}(g))$, define the map
\begin{eqnarray*}
\bar{\nabla} : {\cal X}(G)\times\cal{E}&\longrightarrow&\cal{E}\quad
\hbox{via}\\
(\bar{\nabla}_X\sigma)(g)&=&T_g\sigma\cdot X(g) +
\gamma_g(\bar{X}(g),\sigma (g)).
\end{eqnarray*}
 It is straightforward to check that $\bar{\nabla }$ is a vector bundle 
connection. Therefore, a bilinear map $\gamma_g$ defines the vector
bundle connection \mbox{$\bar{\nabla}:{\cal X}(G)\times\cal{E}
\rightarrow\cal{E}$}. Moreover when $\gamma_g$ is defined as in
 (\ref{gamma}), the push-forward by $\rho$ of the Levi-Civita
 connection is equal to
\begin{eqnarray*}
\rho (\nabla_XY(g)) & = & (g,\bar{\nabla}_XY(g)), \quad 
\end{eqnarray*}
and therefore,
\begin{eqnarray*}
\nabla_XY(g) & = & T_eR_g(\bar{\nabla}_X\bar{Y}(g)).
\end{eqnarray*}
\newtheorem{proposition}{Proposition}[section]
\newtheorem{conclusion}{Conclusion}
\begin{conclusion}
If for a given connection $\nabla$ we think of the map
\mbox{$\gamma_g:T_eG\times T_eG\rightarrow T_eG$} as a generalized
Christoffel map of the push-forward connection $\bar{\nabla}$ under
the right-trivialisation map $\rho$, then we have the formula
\begin{eqnarray*}
\nabla_XY(g)=\rho^{-1}(d\bar{Y}(g)\cdot
X(g)+\gamma_g(\bar{X}(g),\bar{Y}(g))).
\end{eqnarray*}
If in addition we restrict ourselves to the Levi-Civita connection
 coming from a given metric on $G$, we have the formula for the spray
using the Christoffel map
\begin{eqnarray*}
S(X(G))= T\rho^{-1}(X(g),\bar{X}(g),-\gamma_g(\bar{X}(g),\bar{Y}(g))).
\end{eqnarray*}
The above two formulas are analogous to the finite-dimensional formulas 
and they are globally defined on $G$.
\end{conclusion}

Notice we have not used the right-invariance condition, these 
conclusions are true for any metric on $G$.
However, our results below will depend heavily on the right-invariance 
of the metric.
\begin{proposition}\label{prop}
Let $G$ be a $C^1$-manifold which is a topological group with $C^1$
right translation. Suppose that $G$ admits a right-invariant metric.
Then the spray of the corresponding Lagrangian $L(v)=\frac{1}{2}\left\langle 
v,v\right\rangle _g$  is given by
\begin{eqnarray}
S(v)&=&T\rho^{-1}\circ\bar{S}\circ\rho (v)\; ,\;\hbox{where}\nonumber
\\
\bar{S}(g,\xi )\; &=&\; (T_eR_g\cdot\xi ,\xi ,-B(\xi ,\xi
))\quad\hbox{and}\label{ss}\\
B : T_eG&\times&T_eG\rightarrow T_eG \;\hbox{is defined 
implicitly by}\nonumber \\
\left\langle B (\zeta ,\xi ),\eta \right\rangle &=&\left\langle \zeta ,[\xi 
,\eta ]\right\rangle \quad\hbox{for  } \xi ,\eta ,\zeta\in T_eG.\label{B}
\end{eqnarray}
\end{proposition}
{\em Remark}. For the case of Lie groups, the proof of this result can 
be found in Ref.~\onlinecite{AM}.  The operator $B$  was introduced by 
Arnold~\cite{Arr} in Appendix 2.  The above proposition is more general 
as it covers diffeomorphism groups which are of a great interest in 
hydrodynamics.

{\it Proof.}  To verify (\ref{ss})  we need to calculate the
Christoffel map $\gamma_g(\xi ,\xi )$ in (\ref{Sbar2}).

Since $\rho$ is a diffeomorphism, we can push-forward the symplectic
form $\omega_L$ on $TG$ to define the symplectic form
$\omega^s=\rho_*\omega_L$  (the super-script $s$ stands for ``spatial''
because the right trivialisation gives rise to spatial coordinates in 
applications). It can be checked that the push-forward of the spray
$S$ on $TG$ is the Lagrangian vector field  expressed in space 
coordinates and $\rho_*S=\bar{S}$; consequently, 
\begin{equation}
{\bf i}_{\bar{S}}\omega^s\; =\; {\bf d}(\rho_*E). 
\label{I}
\end{equation}
To calculate the left-hand side recall the following formula 
(see Ref.~\onlinecite{AM}):
\begin{eqnarray*}
\omega^s(g,\xi )((v,\zeta ),(w,\eta ))=&-&\left\langle \zeta 
,T_gR_{g^{-1}}(w)\right\rangle _e + \left\langle \eta 
,T_gR_{g^{-1}}(v)\right\rangle _e \\
&-&\left\langle \xi ,[\; T_gR_{g^{-1}}(v),T_gR_{g^{-1}}(w)\;  
]\right\rangle _e \; . \end{eqnarray*}
By this formula we have that
\begin{eqnarray}
\omega^s(g,\xi )((T_eR_g\cdot\xi ,-\gamma_g(\xi ,\xi )),(w,\eta ))&=&
-\left\langle -\gamma_g(\xi ,\xi 
),T_gR_{g^{-1}}(w)\right\rangle _e\nonumber \\
+\left\langle \eta ,\xi \right\rangle _e&-&\left\langle \xi ,[\; \xi 
,T_gR_{g^{-1}}(w)\;  ]\right\rangle _e \; . \label{J}
\end{eqnarray}

Since the metric is right-invariant, it follows that
\begin{eqnarray*}
E\circ\rho^{-1}(g,\xi )&=& {1\over 2}\left\langle T_eR_g\cdot\xi 
,T_eR_g\cdot\xi \right\rangle _g\\
&=&\frac{1}{2}\left\langle\xi ,\xi \right\rangle _e \; .
\end{eqnarray*}
Therefore the right-hand side of (\ref{I}) is equal to
\begin{equation}
{\bf d}(E\circ\rho^{-1})(g,\xi )\cdot (w,\eta )\; =\; \left\langle\xi ,\eta 
\right\rangle _e\; . \label{K}
\end{equation}
From (\ref{I}),(\ref{J}), and (\ref{K})  we may conclude that the value of
$\gamma_g(\xi ,\xi )$ does not depend on the base point $g$. Moreover,
its value is defined by the following relationship:
\begin{eqnarray*}
\left\langle\gamma (\xi ,\xi ),\zeta \right\rangle _e\; =\; \left\langle \xi 
,[\xi ,\zeta ]\; \right\rangle _e\quad \hbox{for}\; \xi ,\zeta\in T_eG\; .
\end{eqnarray*} 
From the definition (\ref{B}) of the operator $B$ it follows that 
$\gamma (\xi ,\xi )=B(\xi ,\xi )$, and hence (\ref{ss}) is true. $\quad \Box $

It is known that 
$-B(\xi ,\xi )=(\nabla_{X_{\xi}}X_{\xi})(e)$ , where
$X_{\xi}(g)=T_eR_g\cdot\xi$ (see Arnold~\cite{Ar}, Bao and Ratiu~\cite{BR}).
Thus, for right invariant vector fields, we also have that
\begin{eqnarray*}
\bar{S}(g,\xi )=(X_{\xi}(g),\xi ,(\nabla_{X_{\xi}}X_{\xi})(e)).
\end{eqnarray*}
\begin{conclusion}
Given a right-invariant metric on $G$, we can find its geodesic
equations by finding the spray. The above formulas show that the spray
is completely defined by either the operator $B$ or the value of the
Levi-Civita connection at the identity.
\end{conclusion}
See remark in Section~\ref{talk}.

\subsection{The Euler-Poincar\'{e} Equations}\label{EP}

The Euler-Poincar\'{e}-Arnold equations are the fundamental result about 
geodesic flow on an arbitrary 
Lie group. See, for example, Theorem 13.8.3 in Marsden and Ratiu~\cite{MR} 
or Appendix 2 in Arnold~\cite{Arr}.  Herein, this 
result is proven for diffeomorphism groups, the configuration space for 
ideal fluid dynamics. The idea of studying geodesics on diffeomorphism
groups in order to do hydrodynamics is due to Arnold~\cite{Ar}.

In order to establish our notation, let us recall some results from 
Refs.~\onlinecite{EM} and~\onlinecite{MEF}.
For $s>\frac{n}{2}$ and $M$ a compact manifold without boundary, 
we may define the Sobolev $H^s $ maps from $M$ into $M$. Let ${\cal D}^s(M) = 
\{\eta \in H^s(M,M)\, |\, \eta \,\,\hbox{is bijective and}\,\, \eta ^{-1} 
\in H^s(M,M)\}$.
If $s>\frac{n}{2} + 1$,  then ${\cal D}^s(M)$ is open in $H^s(M,M)$  and 
hence
is a manifold, but note 
that ${\cal D}^s$ is not a Lie group. but rather a topological group.  
However, like a Lie group, ${\cal D}^s$ has an exponential map which 
associates to every tangent
vector at the identity a one parameter subgroup of ${\cal D}^s$. Such a 
tangent vector is an $H^s$ vector field on $M$ and the one parameter subgroup
is its flow.
If $\pi \colon TM \rightarrow M$ is the canonical projection, one forms 
the Hilbert space
\begin{eqnarray*}
T_{\eta }{\cal D}^s = \{ V:M\rightarrow TM \,|\, V \,\,\hbox{is}\,\, H^s 
\,\,  
\hbox{and} \,\,\pi\circ V = \eta \}\, ,
\end{eqnarray*}
the tangent space at $\eta \in {\cal D}^s$. 
An element $V$ of the tangent space at $\eta \in {\cal D}^s$ is called a 
vector space over $\eta $.

\newtheorem{theorem}{Theorem}[section]
\begin{theorem}\label{Arnold}
Assume that ${\cal D}^s(M)$ is equipped with 
a metric $\left\langle\cdot\; ,\;\cdot 
\right\rangle $ that is invariant under right translations.
Then a curve  $t\rightarrow\eta (t)$  in ${\cal D}^s$  is a geodesic of
this metric if and only $u(t)=T_{\eta(t)}R_{\eta(t)^{-1}}\dot{\eta}(t)
=\dot{\eta}(t)\circ\eta^{-1}(t)$  satisfies
\begin{equation}
{du\over dt}\; =\; -B(u,u),
\label{Euler-Poincare}
\end{equation}
where the operation $B:{\cal X}(M)\times {\cal X}
(M)\longrightarrow{\cal X}(M)$
is defined by the identity
\begin{eqnarray*}
\left\langle B (w,u ),v \right\rangle = 
\left\langle w,[u,v]\right\rangle \quad\hbox{for  } 
u,v,w\in {\cal X}(M). \end{eqnarray*} 
\end{theorem}
{\it Proof} . By proposition (\ref{prop}) the spray $S:T{\cal D}^s\rightarrow
 T^2{\cal D}^s$ is given by
\begin{equation}
S(V)=T\rho^{-1}(V,V\circ\eta^{-1},-B(V\circ\eta^{-1},V\circ\eta^{-1}))
\quad\hbox{for  }V\in T_{\eta}{\cal D}^s.
\label{sss}
\end{equation}
Using the notation adopted from the theory of Lie groups
$G={\cal D}^s\; ,\; {\cal G}={\cal X}(M)$ let us compute $T\rho^{-1}$.
\begin{eqnarray*}
\rho^{-1}:G\times\cal{G}&\longrightarrow& TG \\
 (\eta ,u)&\longmapsto& u\circ\eta \\
T_{(\eta ,u)}\rho^{-1}: T_{\eta}{\cal D}^s\times\cal{G}&\longrightarrow&
T_{u\circ\eta}(TG)\\
\end{eqnarray*}
Let $(\eta (t),u(t))$ be a curve in ${\cal D}^s\times{\cal X} (M)$ such that 
$(\eta (0),u(0))=(\eta ,u)$  and  $(\dot{\eta}(0),\dot{u}(0))=(V,w)$,
then
\begin{eqnarray*}
T_{(\eta ,u)}\rho^{-1}\cdot (V,w)&=& {{d\over dt}\arrowvert}_{t=0}
\rho^{-1}(\eta (t),u(t)) \\
&=&{{d\over dt}\arrowvert}_{t=0}u(t)\circ\eta (t) \\
&=&w\circ\eta\;  + \; Tu\circ V.
\end{eqnarray*} 
Therefore,  
\begin{eqnarray}
S(V)&=&-B(V\circ\eta^{-1},V\circ\eta^{-1})\circ\eta\; +\; 
T(V\circ\eta^{-1})\circ V,\quad\hbox{or}\nonumber \\
S(V)\circ\eta^{-1}&=&-B(V\circ\eta^{-1},V\circ\eta^{-1})\; +\; 
T(V\circ\eta^{-1})\circ (V\circ\eta^{-1}). \label{one}
\end{eqnarray}
For an integral curve $V_t\in T_{\eta (t)}{\cal D}^s(M)$ ,
its pullback $u_t=V_t\circ\eta_t^{-1}$ is a curve in $T_e{\cal D}^s(M)$
that consists of $H^s$-vector fields on $M$.
Then the spray equation,
\begin{equation}
{dV_t\over dt}=S(V_t),
\label{sprayeqn}
\end{equation}
is equivalent to
\begin{eqnarray}
{d\over dt}(u_t\circ\eta_t)&=&S(V_t)\; ,\; \hbox{or}\nonumber\\
{du_t\over dt}&=&S(V_t)\circ\eta_t^{-1}\; -\; Tu_t\circ u_t\label{two}\\
&=&-B(u_t,u_t) + Tu_t\circ u_t-Tu_t\circ u_t\nonumber\\
&=&-B(u_t,u_t).\quad \nonumber
\end{eqnarray} 
Note that the existence of the smooth spray defined by (\ref{sss}) 
follows from agruments in Theorem 3.3 in Shkoller~\cite{SH} as well as
Theorem 4.2 in Ref. \onlinecite{HKMRS}.
See the remark in Section~\ref{talk}.  $ \Box$

\section{ Camassa-Holm equation as a geodesic flow.} \label{geo}
Henceforth, the subscript $t$ will denote the partial 
derivative with respect to $t$. 
We shall apply the general results obtained in section~\ref{main}
to the CH-equation (\ref{CH}). 
For periodic boundary conditions, the configuration space 
is $G={\cal D}^s(S^1)$ with Lie
algebra ${\cal G}={\cal X}^s(S^1)$.  One may also consider (\ref{CH}) on
${\bf R}$ with the appropriate decay conditions at 
infinity guaranteed by the Sobolev class $H^s$.
Formal computations are identical in both cases and, hence, may be 
treated simultaneously.
Consider the $H^1$ metric on 
${\cal G}$ given at the identity by 
\begin{eqnarray*} \left\langle u,v\right\rangle _1=\int 
(uv+u_xv_x)dx\, , \end{eqnarray*}
where the integral may either be taken over $S^1$ or ${\bf R}$.
Camassa and Holm~\cite{CHH} have shown that the Lagrangian for the 
CH-equation (\ref{CH}) is given by the $H^1$ norm (\ref{LagrangianCH}).

Given a metric on a Lie algebra, we can define a metric on all 
\mbox{$T{\cal D}^s$} by right translation so that for $V,W\in T_{\eta}
{\cal D}^s$, 
\begin{eqnarray}
\left\langle V,W\right\rangle 
_{\eta}&=&\left\langle T_{\eta}R_{\eta^{-1}}\cdot
V,T_{\eta}R_{\eta^{-1}}\cdot W\right\rangle _{id}\nonumber \\
&=&\left\langle V\circ\eta^{-1},W\circ\eta^{-1}\right\rangle _1\nonumber \\
&=&\int [(V\circ\eta^{-1})(W\circ\eta^{-1})+(V\circ\eta^{-1})_x
(W\circ\eta^{-1})_x]dx.
\label{metric}
\end{eqnarray}
This metric is right-invariant by definition and it defines the extended 
Lagrangian
\begin{eqnarray*}
L(V)=\frac{1}{2}\left\langle V,V\right\rangle _{\eta}.
\end{eqnarray*}
The main result of this section is
\begin{theorem}
Let $t\rightarrow\eta (t)$ be a curve in the diffeomorphism group
${\cal D}^s$ starting at the identity. Then $\eta (t)$ is 
a geodesic of the metric (\ref{metric})  if and only if the time-
dependent vector field $u(t)=\dot{\eta}(t)\circ\eta^{-1}(t)$
satisfies the CH-equation (\ref{CH}).
\end{theorem}
{\it Proof.}  By the theorem (\ref{Arnold}), the geodesic equations 
for the metric (\ref{metric}) are equivalent to (\ref{Euler-Poincare}). 
From the definition of the operator $B$, we have that
\begin{eqnarray*}
\left\langle B (w,u ),v \right\rangle = \left\langle 
w,[u,v]\right\rangle &=&\int (-uv_x+u_xv)w+(-uv_x+u_xv)_xw_x\; 
dx\\ &=&\int (u_xw+(uw)_x)v-w_{xx}(-uv_x+u_xv)\; dx\\
&=&\int (2u_xw-2u_xw_{xx}+uw_x-uw_{xxx})\; v\; dx\\
&=&\int (2u_x(1-\partial^2_x)w+u(1-\partial^2)w_x)\; v\; dx.
\end{eqnarray*}
Furthermore, 
\begin{eqnarray*}
\left\langle B (w,u ),v \right\rangle &=&\int (B (w,u )v+B (w,u )_xv_x)\; 
dx\\ &=&\int ((1-\partial^2)B (w,u ))\; v\; dx,
\end{eqnarray*}
and hence the formula for the operator $B$ is
\begin{equation}
B(w,u)=(1-\partial^2)^{-1}(2u_x(1-\partial^2)w+u(1-\partial^2)w_x).
\end{equation}
Now we obtain the Euler-Poincar\'{e}-Arnold equation:
\begin{eqnarray*}
\frac{\partial u}{\partial t}&=&-B(u, u) \\
&=&-(1-\partial^2)^{-1}(2u_xu+uu_x-2u_xu_{xx}-uu_{xxx})\\
&=&-(1-\partial^2)^{-1}(3uu_x-2u_xu_{xx}-uu_{xxx}).
\end{eqnarray*}
This completes the proof that the geodesic equations for the metric
coming from the $H^1$~ inner product on the Lie algebra of vector fields 
${\cal G}$ are equivalent to the CH-equation~(\ref{CH}).

{\it Remark.} The Lie-algebra bracket $[u,v]$ on ${\cal X}(M)$
is minus the Jacobi-Lie bracket (for an explanation refer to Marsden
and Ratiu~\cite{MR}, Chapter 9). $ \Box$

\paragraph{Alternative derivation.}
Below, we begin to compute the geodesic equations for the 
metric (\ref{metric}) by calculating the spray of the corresponding 
Lagrangian.  Camassa and Holm~\cite{CHH} have shown that the
CH-equation can be expressed in the integral form
\begin{eqnarray*}
u_t+uu_x&=&-(1-\partial^2)^{-1}\partial (u^2+\frac{1}{2}u_x^2)\\
&=&-\int e^{-|x-y|}(uu_y+\frac{1}{2}u_yu_{yy})\; dy.
\end{eqnarray*}
Equation (\ref{one}) together with the fact that $Tu\circ u$ is simply
$uu_x$ in one dimension shows that the spray is equal to
\begin{equation}
S(V)=-(1-\partial^2)^{-1}\partial ((V\circ\eta^{-1})^2+\frac{1}{2}
(V\circ\eta^{-1})^2_x)\circ\eta. \label{sprayCH}
\end{equation}
Letting $u=V\circ\eta^{-1}$ verifies the claim.

\section{Discussion} \label{talk}
We would like to emphasize that we built a right invariant metric on 
${\cal D}^s $ by taking the $H^1$   inner product on the tangent space at 
the 
identity and right-translating it over the whole space. This does not
coincide with the {\it usual} $H^1$ metric on each fiber $T_\eta {\cal 
D}^s$; see the remark after Theorem 4.1 in Ref.~\onlinecite{HKMRS}.
To illustrate the difference of two approaches let us compare the geodesics
of the $L^2$ metric with the right-invariant $L^2$ metric in the 
one-dimensional case. 

 A curve $\eta (t)\,\in\, {\cal D}^s$ is a geodesic of the 
$L^2$-metric if and only if the corresponding spatial velocity field 
\[
u=V\circ\eta^{-1} 
\]
satisfies Burger's equation:
\begin{equation}
u_t\ +\ uu_x=0.
\end{equation}
The corresponding Euler-Lagrange equations for the material velocity 
$V=\dot{\eta}$ are given by
\[
V_t=0\, .
\]
The spray of this metric is equal to zero and hence smooth; however, as the 
metric is not right invariant, the Euler-
Poincar\'{e} theorem does not apply.

For the right-invariant $L^2$ metric the Euler-Poincar\'{e} equations are
given by 
\begin{equation}
u_t\, +\, 3uu_x=0\, .
\end{equation}
The corresponding Euler-Lagrange equations are given by 
\begin{eqnarray*}
\eta_X\dot{V}\  +\ 2VV_X=0\, , 
\end{eqnarray*}
where $\eta_X$ is the Jacobian of $\eta$, and $X$ denotes the material coordinate
of the fluid particle.
The spray in this case is given by
\begin{equation}
S(\eta ,V)=-\frac{2}{\eta_X}\ VV_X.
\end{equation}
Since there is a loss of derivatives, the spray is not smooth (c.f. Remark 3.5 in Shkoller~\cite{SH}). 

As we see from the above calculations, the two equations in the spatial 
velocities differ only in a scalar coefficient multiplying the 
derivative term, however, the corresponding sprays are completely different.

{\em Remark}. We note that equation (\ref{sprayCH}) for the geodesic spray
of the right-invariant $H^1$ metric on either $S^1$ or ${\bf R}$ has no
derivative loss and hence shows that the CH-equation is an ordinary 
differential equation on the group ${\cal D}^s$.  Thus, existence and 
uniqueness of solutions to (\ref{CH}) may be obtained by standard Picard 
iteration argument in the event that $S$ is locally Lipschitz.

Lemmas 3.1 and 3.2 of Shkoller~\cite{SH} show that $S$ is $C^1$, and 
hence the result follows.  See Refs. \onlinecite{HKMRS} and \onlinecite{SH}
for the well-posedness of the geodesic flow of the diffeomorphism groups 
on $n$-dimensional  Riemannian manifolds.

It would be interesting to study the Lagrangian stability of the CH-equation,
and this requires analysis of the curvature operator.  Misiolek~\cite{M} has
computed the sectional curvature of ${\cal D}^s(S^1)$.  Shkoller~\cite{SH} has
obtained an explicit form for the $H^1$ covariant derivative 
$\tilde{\nabla}^1$ on volume preserving diffeomorphism groups and has 
proved that the weak curvature tensor of $\tilde{\nabla}^1$ is a bounded 
trilinear operator in the $H^s$ topology.  We would like to explore these
type of estimates on the full diffeomorphism group of the circle, as well
as investigate the role of generalized flows in peakon dynamics.     

\section{Acknowledgments} \label{thanks}
I am thankful to Darryl D. Holm and Tudor Ratiu for their kind
guidance during the course of this work. Their helpful suggestions,
patient discussions, and sincere interest in the subject helped me 
to build confidence in this work. I also would like to thank 
Jerrold E. Marsden and Steve Shkoller for their comments on the
paper.  Their comments helped to make the paper more clear and
detailed.

\end{document}